\documentclass[aps,prl,reprint,superscriptaddress]{revtex4-2}

\usepackage{graphicx}
\usepackage{dcolumn}
\usepackage{bm}

\usepackage{xr}

\usepackage[utf8]{inputenc}
\usepackage{amsmath,amssymb,amsfonts}
\usepackage{graphicx}
\usepackage{dcolumn}
\usepackage{bm}
\usepackage{booktabs}
\usepackage{physics}
\usepackage{csquotes}
\usepackage{comment}
\usepackage{graphicx}
\usepackage[table]{xcolor}
\usepackage{tabularx,colortbl}
\usepackage{soul}
\usepackage{caption}
\usepackage[singlelinecheck=false]{subcaption}
\usepackage{siunitx}
\usepackage{lipsum}
\usepackage[ruled,lined]{algorithm2e}
\usepackage{mathtools}
\usepackage{siunitx}
\usepackage{textcomp}
\usepackage{svg}
\usepackage{soul}
\usepackage{xpunctuate}
\usepackage{soul}
\usepackage{subfiles}
\usepackage{hyperref}
\usepackage[capitalise]{cleveref}

\SetCommentSty{mycommfont}

\crefname{algorithm}{Alg.}{Algs.}
\crefname{appendix}{App.}{Apps.}

\newcommand{\mcite}[1]{Ref.~\cite{#1}}
\newcommand{\mcitep}[1]{Refs.~\cite{#1}}

\makeatletter
\newcommand*{\addFileDependency}[1]{
        \typeout{(#1)}
        \IfFileExists{#1}{}{\typeout{No file #1.}}
}
\makeatother

\newcommand*{\myexternaldocument}[1]{%
        \externaldocument{#1}%
        \addFileDependency{#1.tex}%
        \addFileDependency{#1.aux}%
}

\myexternaldocument{sm}

\newcommand*{\oh}[1]{\hat{#1}}
\newcommand*{\Htwo}{H\textsubscript{2}}
\newcommand*{\Ry}{R\textsubscript{y}}
\newcommand*{\Rz}{R\textsubscript{z}}

\newcommand{\ca}{c.a\xperiod}
\newcommand{\ibmqjakarta}{\texttt{ibmq\textunderscore{}jakarta}}
\newcommand{\ibmqosaka}{\texttt{ibmq\textunderscore{}osaka}}

\DeclareSIUnit{\hartree}{Ha}
\DeclareSIUnit{\dt}{dt}
\DeclareSIUnit{\pibar}{\ensuremath{2\pi}}
\DeclareSIUnit{\ebar}{\ensuremath{\hbar}}

\newcommand{\nd}{{\vphantom{\dagger}}}
\begin{document}

\preprint{APS/123-QED}

\title{Molecular groundstate determination via short pulses on superconducting qubits}

\author{Noga Entin}
\thanks{These authors contributed equally to this work.}
\affiliation{The Engineering Faculty, Bar-Ilan University, Ramat-Gan 52900, Israel}
\affiliation{Department of Mathematics, Bar-Ilan University, Ramat-Gan 52900, Israel}
\affiliation{Center for Quantum Entanglement Science and Technology, Bar-Ilan University, Ramat-Gan 52900, Israel}

\author{Mor M. Roses}
\thanks{These authors contributed equally to this work.}
\affiliation{Department of Physics, Bar-Ilan University, Ramat-Gan 52900, Israel}
\affiliation{Center for Quantum Entanglement Science and Technology, Bar-Ilan University, Ramat-Gan 52900, Israel}

\author{Reuven Cohen}
\affiliation{Department of Mathematics, Bar-Ilan University, Ramat-Gan 52900, Israel}

\author{Nadav Katz}
\affiliation{Racah Institute of Physics, the Hebrew University of Jerusalem, Jerusalem 91904, Israel}

\author{Adi Makmal}
\email{adi.makmal@biu.ac.il}
\affiliation{The Engineering Faculty, Bar-Ilan University, Ramat-Gan 52900, Israel}
\affiliation{Center for Quantum Entanglement Science and Technology, Bar-Ilan University, Ramat-Gan 52900, Israel}

\date{\today}

\newcommand\mor[1]{\textcolor{blue}{[Mor: #1]}}
\newcommand\noga[1]{\textcolor{red}{[Noga: #1]}}
\newcommand\adi[1]{\textcolor{purple}{[Adi: #1]}}
\newcommand\checkit[1]{\textcolor{orange}{[#1]}}
\newcommand\finetune[1]{\textcolor{purple}{#1}}
\newcommand\finetuned[1]{\textcolor{blue}{#1}}
\newcommand{\replace}[1]{\textcolor{green}{Replace?: [#1]}}

\begin{abstract}
    Quantum computing is currently hindered by hardware noise. We present a freestyle superconducting pulse optimization method, incorporating two-qubit channels, which enhances flexibility, execution speed, and noise resilience.
    A minimal 0.22 ns pulse is shown to determine the \Htwo{} groundstate to within chemical accuracy upon real-hardware, approaching the quantum speed limit. Similarly, a pulse significantly shorter than circuit-based counterparts is found for the LiH molecule, attaining state-of-the-art accuracy. The method is general and can potentially accelerate performance across various quantum computing components and hardware.
\end{abstract}


\maketitle
\section{Introduction}
Current quantum devices suffer from significant hardware errors,
restricting their ability to fulfill the promise of quantum computing \cite{Preskill_2018,Bharti_2022}.
Variational quantum algorithms (VQAs) have become popular due to their shallow circuits \cite{cerezo2021variational,peruzzo2014variational_org_vqe,farhi2014quantum,bravo2023variational,mangini2023variational}, but despite much scientific effort, they are still hampered by hardware noise \cite{cerezo2021variational}. To improve their noise resilience, one strategy is to develop new, shallower, quantum circuit architecture designs \cite{Kandala_2017, Grimsley_2019, ostaszewski2021reinforcement,he2023gnn, ding2023multi}. An alternative approach has been recently attempted via direct manipulation of the hardware degrees of freedom.

In superconducting qubits, each quantum gate is implemented via a fixed set of microwave pulses \cite{gustavsson2013improving}.
Quantum optimal control (QOC) techniques, where pulse shapes are carefully designed \cite{koch2022quantum}, e.g.\ gradient ascent pulse engineering (GRAPE) \cite{khaneja2005optimal}, can be used to improve the fidelity of the gates and speedup the circuits’ execution times \cite{wu2018data,zong2021optimization}.
Such pulse optimization methods can also be utilized for state-preparation, as demonstrated for example in \mcite{PhysRevLett.110.100404} and very recently for two-qubit systems with bounded amplitude, see \mcite{li2023optimal} and references therein.
Machine learning techniques have further contributed to the improvement of precise state-preparation via pulse engineering methods \cite{an2019deep,porotti2023gradient}.

Variational pulse optimization aims at the minimization of a predefined VQA cost-function, rather than the precise \textit{fidelity} of a target gate or state.
\citeauthor{PRXQuantum.2.010101} have initially explored conceptual connections between QOC and VQAs on a theoretical level \cite{PRXQuantum.2.010101}. Later on,
various concrete strategies emerged, which we categorize into three groups:
\textbf{(a) The gate-based} approach leverages conventional quantum gates, e.g.\ by transpiling gated-circuit Ansatz into corresponding standard DRAG (derivative removal by adiabatic gate) pulses and optimizing their amplitudes and frequency envelopes \cite{liang2024napa,9951187,liang2022hybrid};
\textbf{(b) The Shape-based} approach, builds on familiar pulse shapes, such as Gaussian and DRAG pulses, and tweaks them for single or two-qubit operations \cite{meirom2023pansatz, egger2023study,pan2023experimental}.
\textbf{(c) The square pulse method} takes a more straightforward approach with discrete time square pulses, adjusting merely the amplitudes, as introduced by \citeauthor{meitei2021gate} 
\cite{meitei2021gate} for superconducting qubits.
A similar protocol was then demonstrated via classical simulations on Rydberg atoms hardware \cite{de2023pulse}.
Notably, the study in \mcite{meitei2021gate} manipulated merely single qubit's degrees of freedom without controlling the two-qubit channels.

We introduce a fully flexible scheme for superconducting devices that combines the square-pulse method \cite{meitei2021gate} with direct, bi-directional control over two-qubit channels for enhancing flexibility and expressibility.
Inspired by machine learning's ability to learn ``without being explicitly programmed" \cite{samuel1959some}, our method
autonomously discovers the most effective pulses within the device's physical constraints, by tuning the pulse amplitudes across both single and two-qubit channels, at each discrete time step.

We tested our method on groundstate calculations of the \Htwo{} and LiH molecules and demonstrated a significant reduction in pulse duration, up to three orders of magnitude, and substantially improved accuracy compared to previous gate-based and pulse-based results on real-hardware. The reduction in pulse duration was achieved in two steps: first, we employed the qubit-efficient variational quantum selected-configuration-interaction (VQ-SCI) scheme \cite{yoffe2023qubitefficient}, which finds molecular groundstates using relatively few qubits and shallow circuits; and second, we replaced the VQ-SCI gated Ansatz with our \textquote{freestyle} pulse, thereby reducing the pulse duration further.
Notably, the sufficiency of our exceptionally short pulses aligns with the \enquote{bang-bang} control theory principles, advocating for abrupt and robust control applications for quick and efficient outcomes \cite{yang2017optimizing,innocenti2020ultrafast,asthana2023leakage}.

This letter begins with a background on pulse design and the VQ-SCI algorithm, 
followed by a description of the freestyle pulse optimization scheme. We then evaluate its performance on the \Htwo{} molecule (1 qubit), upon IBMQ hardware, attaining pulse duration that closely reaches the theoretical quantum speed limit (QSL).
This section also addresses leakage and measurement errors. Subsequently, the scheme is applied to the groundstate determination of the LiH molecule (3 qubits). Unparalleled precision is achieved with a pulse duration that is six times shorter than that of the counterpart circuit duration. Finally, we outline future research directions.

\section{Background}\label{sec:background}
\subsection{The pulse Hamiltonian in superconducting qubits}
In superconducting setting, the Hamiltonian over $N_q$ qubits is given by (see Supplementary Material) \cite{blais2021circuit}: 
\begin{align}
\label{eq:Hfull}
    \oh H_\mathrm{full}=&
    \sum\limits_k^{N_q}\bqty{
    \oh H_\mathrm{Drive}^{(k)}
    +
    \sum\limits_{l\in\mathcal{N}_k}\pqty{
    \oh H_\mathrm{Control}^{(k,l)}
    +\oh H_\mathrm{CR}^{(k,l)}
    }
    },
\end{align}
where $\oh H_\mathrm{Drive}^{(k)}$ is a tunable drive channel for the $k$'th qubit,
$\oh H_\mathrm{Control}^{(k,l)}$ is a tunable two-qubit control channel of qubit $k$ over qubit $l$, $\oh H_\mathrm{CR}^{(k,l)}$ is the fixed cross-resonance Hamiltonian, which generates an uncontrolled interaction between qubits $k$ and $l$, and  $\mathcal{N}_k$ accounts for all qubits connected to the $k$'th qubit in the device topology.

The single qubit drive channels $\oh H_\mathrm{Drive}^{(k)}$ and the two-qubit control channels $\oh H_\mathrm{Control}^{(k,l)}$ can be tuned via user-defined time-dependent complex functions $d_k\pqty{t}$ and $u_{k,l}\pqty{t}$, respectively, whose magnitude is restricted to $\abs{d_k\pqty{t}}^2\le1$, and $\abs{u_{k,l}\pqty{t}}^2\le1$, to ensure physical limits.

\subsection{The variational quantum selected configuration-interaction (VQ-SCI) algorithm}
The variational quantum eigensolver (VQE) algorithm marked the beginning of VQAs dedicated to finding the groundstate energy of chemical systems \cite{peruzzo2014variational_org_vqe}.
Recently, a qubit-efficient alternative algorithm was proposed, termed variational quantum selected configuration interaction (VQ-SCI), which serves the same purpose but with fewer qubits, by adopting a different encoding scheme \cite{yoffe2023qubitefficient}:
it is based on first, rather than second, quantization, where the computation is carried out on the basis of Slater determinants (or configurations). Similar to classical selected configuration interaction (SCI) schemes, and contrary to the full configuration interaction (FCI) method, which accounts for all possible Slater determinants,
the VQ-SCI is based on selecting only the most significant ones.
This reduces computational burden with a controlled compromise in accuracy \cite{tubman2020modern}.

We demonstrated our pulse scheme through the VQ-SCI method for two reasons: First,  VQ-SCI is less susceptible to noise, due to the qubit reduction. For the \Htwo{} and LiH molecules it requires just 1 and 3 qubits, respectively, instead of 2 and 4 qubits in VQE,
see \mcite{yoffe2023qubitefficient} and the {Supplementary Material};
Second, VQ-SCI conveniently sets the Hartree-Fock (HF) solution as the all-zero state $\ket{0\ldots0}^{\otimes q}$,
whereas VQE requires applying an extra $X$ gate to as many qubits as the number of active electrons, a straightforward but time-consuming (about 71 ns) and noise-introducing process  \footnote{The Hartree-Fock solution is the closest single-determinant approximation to the true many-body groundstate wavefunction in terms of energy difference.}. Since we opted for short and accurate calculations, these are valuable advantages.

\section{The freestyle pulse method}
\label{sec:our_method}
The central idea of our freestyle pulse optimization scheme, similar to the ctrl-VQE approach proposed in \mcite{meitei2021gate}, is the discretization of the continuous pulse into small time intervals \unit\dt, each with a tunable complex amplitude, that is optimized individually. This discretization grants optimal flexibility in shaping the pulse, without the need to adhere to any predefined function.
However, in contrast to \mcite{meitei2021gate}, our scheme accounts directly for dedicated 2-qubit pulse channels, thereby facilitating natural and direct entanglement generation, being more compatible with real devices. We account for one single-qubit drive channel per qubit and two directed two-qubit control channels per qubit pair, amounting to a total of $q+q(q-1)=q^2$ complex pulse channels, for $q$ qubits in an all-to-all topology.
For $N$ time-bins the number of tunable real parameters is given by $2Nq^2$.

\begin{figure}[h!]
  \centering
  \includegraphics[width=0.55\linewidth]{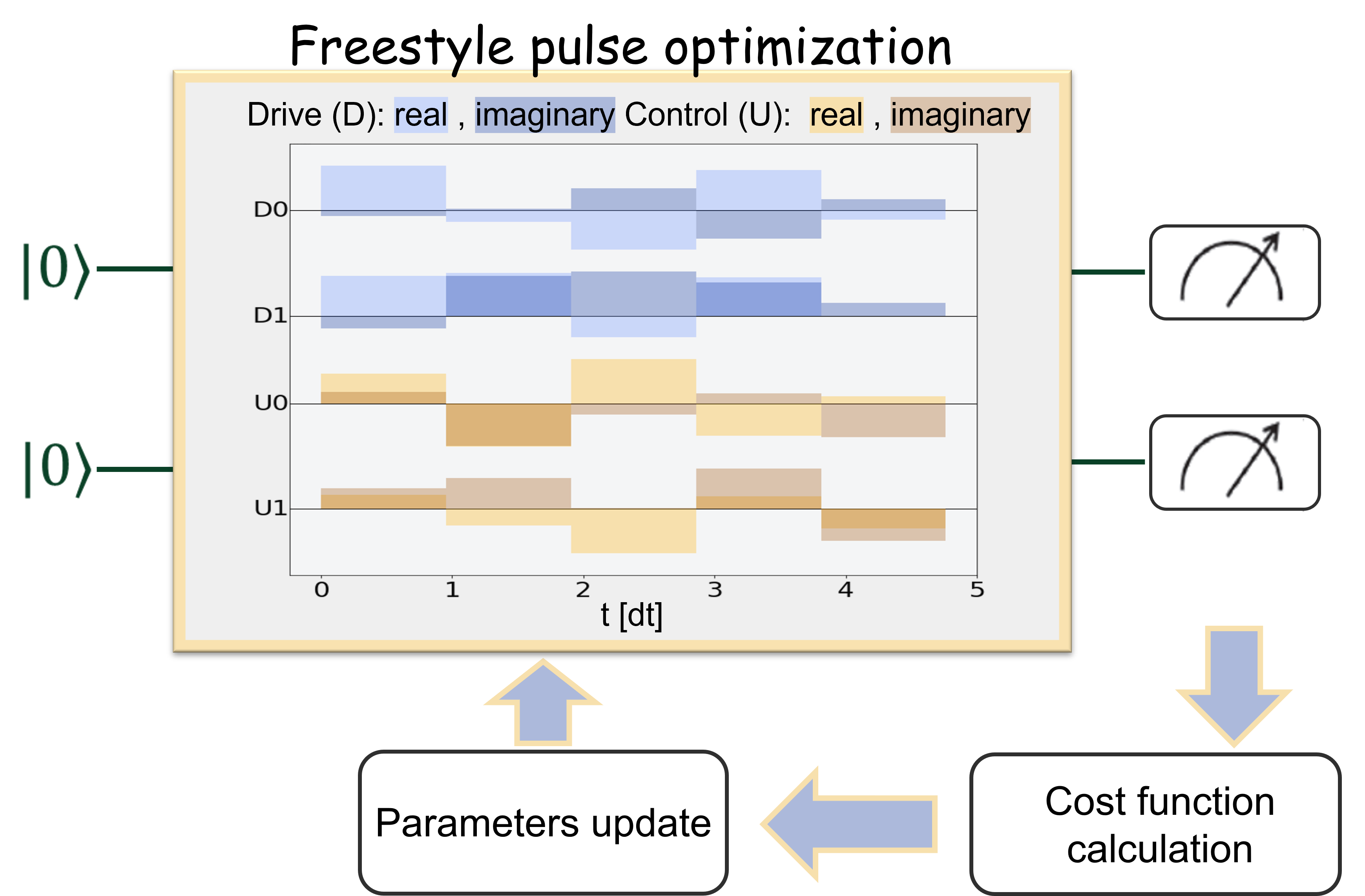}
  \caption{\small Illustrating the freestyle pulse optimization scheme.}
  \label{fig:freestyle_pulse_scheme}
\end{figure}

The scheme is illustrated in
\cref{fig:freestyle_pulse_scheme} for a circuit with $q=2$ qubits and $N=5$ time-bins, and outlined in \cref{alg:pulse}.
It begins with initializing the pulse Hamiltonian of \cref{eq:Hfull}, following which the pulse is executed and the qubits are measured to evaluate the related cost-function; The pulse parameters are then tuned to steer the cost-function towards a global minimum, repeatedly, until convergence.
The pulse duration accounts only for the state preparation (without measurement) and is given by $N$\unit{\dt}, where \unit{\dt} is the time length of each time bin.

\begin{algorithm}
    \DontPrintSemicolon
    \LinesNotNumbered
    \scriptsize
    \caption{Freestyle pulse optimization}
    \label{alg:pulse}
    \SetKwInOut{Input}{input}
    \SetKwInOut{Output}{output}
    \SetKwInOut{Init}{Initialize}
    \KwIn{ $q, N,f_C\pqty*{\va*\theta}$ \tcp*[l]{\#qubits, \#time-bins, cost}}
    \KwOut{$\va*{\theta}^*$, $f_C\pqty*{\va*{\theta}^*}$ \tcp*[l]{optimal parameters and cost}}
    \Init{$\va*\theta$}
    \While{not converged}{
    Run on quantum computer and sample to assess $f_C\pqty*{\va*{\theta}}$ \\
    Update tunable parameters $\va*\theta$ via   classical optimization
    }
    \Return $\Bqty{\va*\theta, f_C\pqty*{\va*\theta}}$
\end{algorithm}

\section{Results}
\label{sec:results}

\subsection{Computational setup}
We illustrate the freestyle pulse optimization on determining the groundstates of \Htwo{} and LiH, within the VQ-SCI framework. The equilibrium interatomic distance is taken to be \SI{0.745}{\angstrom} for \Htwo{} and \SI{1.5}{\angstrom} for LiH.
We conducted noisy simulations and real-hardware experiments on IBMQ devices, utilizing the \textquote{qiskit-dynamics}\cite{qiskitdynamics} package for simulations that reflect the device's connectivity, together with its $T_1$ and $T_2$ values. These simulations included higher energy levels of each Transmon to assess leakage, as detailed in the {Supplementary Material}. We initialized pulses with zero amplitudes, recovering the HF solution, and employed basic readout error mitigation techniques \cite{bravyi2021mitigating} to improve fidelity. To avoid statistical shot-noise, we simulated the system's statevector.
To accommodate IBMQ's hardware constraints on pulse duration, we padded the pulse with zero amplitudes whenever needed 
\footnote{We employed `right' padding by appending zero amplitude pulses to the end of the pulse, effectively delaying its stopping time. This choice showed marginally quicker convergence in simulations compared to the alternative `middle' and `left' padding.}.
Finally, the gradient-free COBYLA optimizer \cite{COBYLA_powell1994direct} was used  \footnote{COBYLA's primar hyperparameter is \emph{rhobeg}, representing the initial step-size. Noisy simulations revealed that for the H$_2$ molecule at interatomic distances up to \SI{1.5}{\angstrom}, an optimal \emph{rhobeg} was 0.05, increasing to 0.1 for larger distances. For LiH at \SI{1.5}{\angstrom}, noiseless simulations indicated an optimal \emph{rhobeg} of 0.8.}.

\subsection{The \Htwo{} molecule}
The 2-qubit Ansätze for \Htwo{} in VQE typically have a pulse duration of hundreds of nanoseconds \cite{liang2024napa}. The VQ-SCI method enables the \Htwo{} groundstate solution to the exact FCI energy value with a single-qubit \Ry{} gate \cite{yoffe2023qubitefficient}, resulting in a pulse duration of
\SI{\approx71}{\nano\second} on IBMQ devices. 
Next, we present noisy simulation and real-hardware VQ-SCI calculations for \Htwo{}  with pulses ranging from 0.22 ns to 2.22 ns on \ibmqjakarta{}, where the minimal time-bin was $\unit\dt=\SI{0.22}{\nano\second}$.
Note that imperfections in experimental equipment result in pulse smearing, effectively extending the duration of a `pulse' beyond the ideal \SI[round-precision=0]{1}{\dt}.

\paragraph{\textbf{Noisy simulations}}

\begin{figure}[h]
  \centering
  \includegraphics[width=0.7\linewidth]{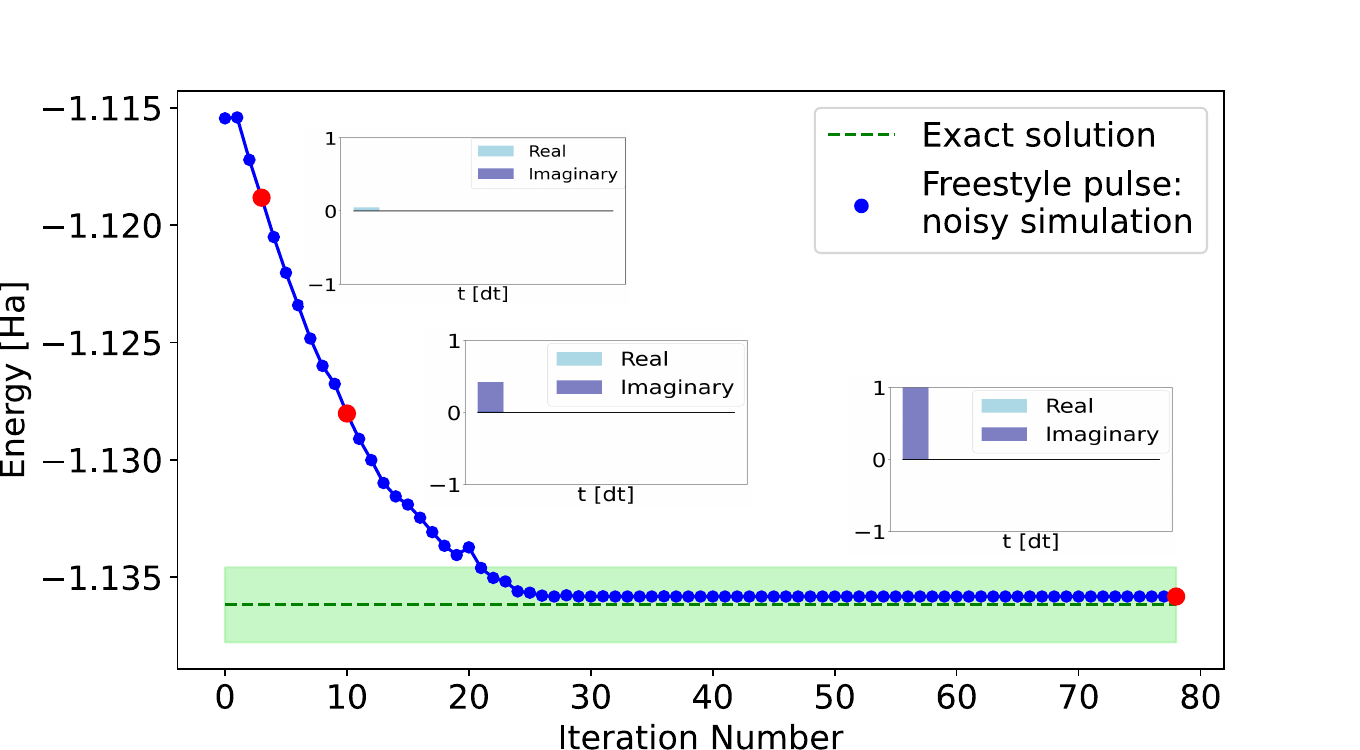}
  \caption{\small\textbf{Noisy simulations} for \textbf{\Htwo{}} at the equilibrium distance 
  with a \SI{0.22}{\nano\second} pulse: the groundstate total energy is shown per iteration (blue). Resulting pulses are depicted at three distinct iterations (red dots). The dashed green line and the green area mark the FCI energy and the c.a. region.}
  \label{fig:convergence noisy simulation}
\end{figure}

\begin{figure}[h]
  \centering
  \includegraphics[width=0.45\linewidth]{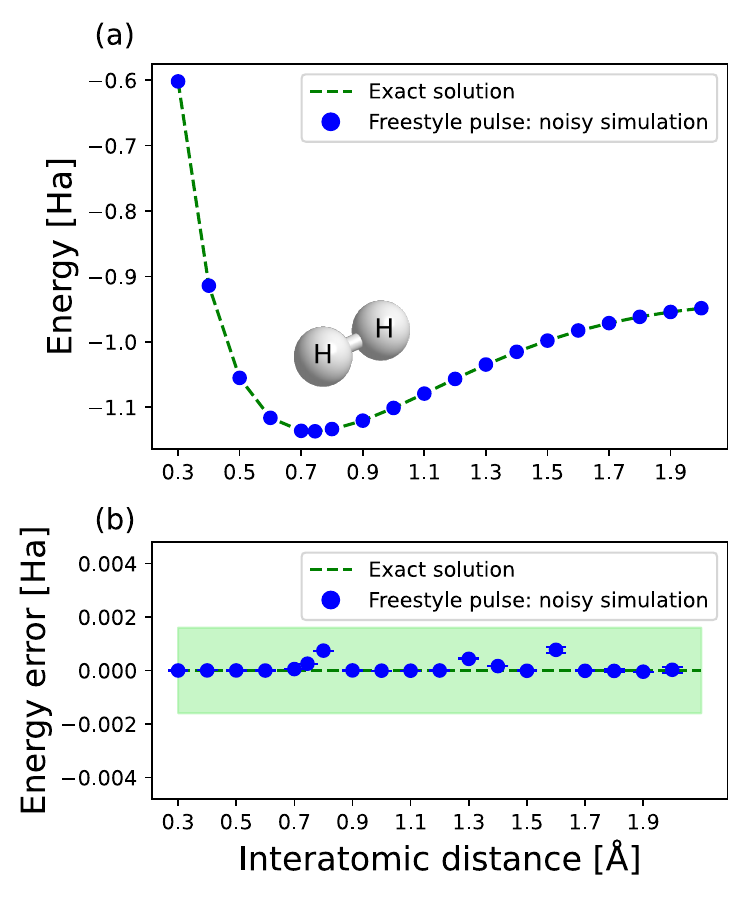}
  \caption{\small \textbf{Noisy simulations} for \textbf{\Htwo{}}: (a) groundstate total energy per interatomic distance; (b) zoom-in view of energy error relative to exact FCI. Error bars present the standard deviation of the last 10 iterations. The dashed green line and the  green area mark the FCI energy and the c.a. region.}
  \label{fig:energy and error dist scheme}
\end{figure}

\Cref{fig:convergence noisy simulation} shows the total groundstate energy of the \Htwo{} molecule at the equilibrium distance
as calculated using the shortest pulse duration of $\SI{1}{\dt}=\SI{0.22}{\nano\second}$ through noisy simulations, plotted against the number of iterations.
It is seen that the iterative process converges rapidly to the \ca region, defined as \SI{\pm.0016}{\hartree} from the FCI energy.
We focus on pulses from three different iterations (red dots). It is seen that only the imaginary part of the pulse's amplitude increases with the iterative process.
This dynamic can be understood within the rotating wave approximation (RWA) frame, see {Supplementary Material}. \Cref{fig:energy and error dist scheme}(a) next depicts the attained \Htwo{} groundstate energy per interatomic distance and \cref{fig:energy and error dist scheme}(b) shows the corresponding deviation from the FCI calculation. The results are shown to be within the \ca region across all distances.

\begin{figure}[h!]
  \centering
  \includegraphics[width=0.5\linewidth]{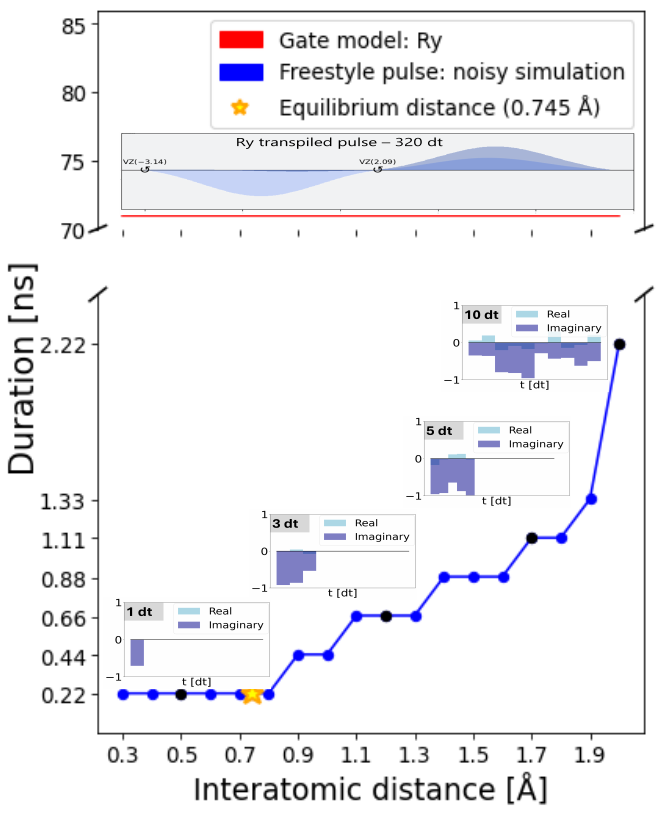}
 \caption{\small \textbf{Noisy simulations} for \textbf{\Htwo{}}: the minimal pulse duration for achieving \ca is plotted per interatomic distance, for the gated model (red) and the freestyle pulse (blue), along the resulting pulses of both methods.}
  \label{fig:minimal pulse len}
\end{figure}

At each interatomic distance, we began with the shortest possible pulse duration (1 \unit{\dt}) and incrementally extended it by one \unit{\dt} at a time until reaching \ca.
\Cref{fig:minimal pulse len} depicts the minimal pulse duration required for the gated circuit and the freestyle pulse method to reach \ca, per interatomic distance.
In the gated circuit, a single \Ry{} gate was used, resulting in a fixed pulse duration of \SI{\approx71}{\nano\second} (red line) regardless of distance \footnote{The  IBMQ circuit transpilation modulates the rotation angle solely through pulse amplitude, without varying its duration.}.
In contrast, the freestyle method requires a maximum of \SI{\approx2.22} {\nano\second}
 \ at a bond length of \SI{2}{\angstrom}, and as little as \SI{1}{\dt}\SI{=0.22}{\nano\second} at equilibrium, showcasing a significant reduction.

\begin{figure}[h!]
  \centering
  \includegraphics[width=0.55\linewidth]{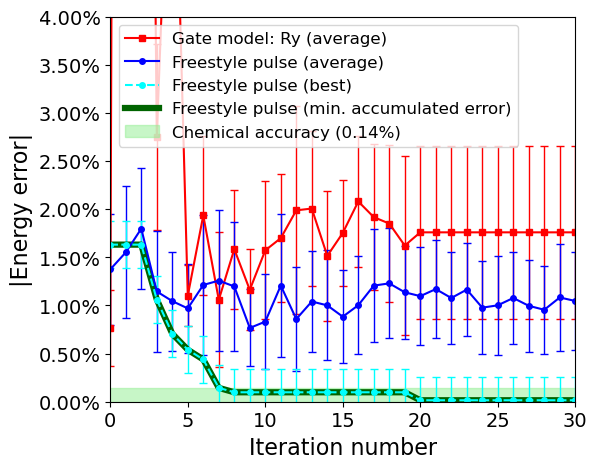}
  \caption{\small \textbf{Real-hardware} (\ibmqjakarta{}) results for \textbf{\Htwo{}} at equilibrium distance. 
  Percentage error deviation from exact FCI energy shown per iteration: averaged over 3 \Ry{} gate experiments (red), averaged over 4 freestyle \SI{0.22}{\nano\second} pulse experiments, and a single freestyle pulse experiment (cyan). The dark green curve depicts the minimal accumulated error at each iteration.
  The light green area marks the \ca region.}
 \label{fig:real-hardware}
\end{figure}

\paragraph{\textbf{Real-hardware}} \label{sec:real-hardware}
We ran the freestyle pulse scheme on the \ibmqjakarta{} device for the \Htwo{} at the equilibrium distance,
using $10^5$ shots.
A single pulse of \SI{1}{\dt}
duration, initialized to zero, was optimized. We ran 4 such experiments.
Our best execution resulted in an FCI deviation of \SI{.2}{\milli\hartree}, reached at the 20th iteration. \Cref{fig:real-hardware} depicts the absolute FCI error deviation $\abs{\pqty{E_\mathrm{iter}-E_\mathrm{FCI}}/E_\mathrm{FCI}}$, per iteration, in percentage. It shows the average of the different freestyle pulse experiments (solid blue), along with the best execution (dashed cyan), shown to converge towards the \ca{} region. This is highlighted by the corresponding minimal accumulated error curve (dark green). The error bars in the average curve present the standard error of the mean (SEM), whereas, for the single execution curve, the error bars correspond to the X-Z measurement errors.

\Cref{fig:real-hardware} further shows the average error of 3 different \Ry{} calculations (solid red). To facilitate a fair comparison,
both the \Ry{} and freestyle pulse experiments were conducted within the VQ-SCI scheme, on the least noisy qubit of the \ibmqjakarta{} device, using identical numbers of shots and error mitigation, with the Ansatz being the only variable. It is seen that the gated \Ry{} experiments converged to a higher average absolute error deviation of about \SI{1.76}{\percent} from the FCI energy. Moreover,  in contrast to the freestyle pulse results, hardly any \Ry{} iteration entered the \ca{} region \footnote{Note that in \mcite{yoffe2023qubitefficient} \ca{} was better achieved. We attribute the degradation in performance to the wear-out of the hardware.}.
The pulse calculations reduced the averaged error significantly to $\approx$\SI{1}{\percent}.

\Cref{table:H2} presents the freestyle pulse duration and the FCI error deviation in comparison to previous real-hardware pulse optimization results carried in Refs.~\cite{liang2024napa} and~\cite{meirom2023pansatz}, where VQE calculations with 2 qubits were performed.
Our optimized pulse duration of \SI{1}{\dt} $\approx$ \SI{0.22}{\nano\second} is more than two orders of magnitude shorter than previously achieved.
Remarkably, a detailed calculation based on Refs.~\cite{giovannetti2003quantum,giovannetti2004speed}, using the exact device parameters, shows that our experimental pulse duration nearly matches the theoretical QSL \cite{deffner2017quantum}. This limit represents the minimal time required for a quantum system's evolution from a given initial state to a desired target state. Although the QSL may be surpassed \cite{atia2017fast}, it provides an order-of-magnitude benchmark for the minimal pulse duration, serving as a valuable figure of merit.

\begin{table}[]
    \scriptsize
    \centering
    \caption{ 
    \small \textbf{Real-hardware} pulse optimization results for \textbf{\Htwo{}} groundstate calculation. Per study, the table specifies: qubits number, utilized method, interatomic distance (i.a.d), pulse duration, and the FCI energy deviation.}
    \label{table:H2}
    \begin{tabular}{
    @{}%
    l%
    c
    c
    S[table-format=1]
    c
    S[table-format=+2.2(2), round-mode=uncertainty, separate-uncertainty]
    @{}%
    }
    \toprule
    &{$\#q$}
    &{alg.}
    &{i.a.d.\ [\unit{\angstrom}]}
    &{T [\unit{\nano\second}]}
    &{$\Delta_\mathrm{FCI}~\bqty{\unit{\milli\hartree}}$}
    \\
    \midrule
    NAPA \cite{liang2024napa} & 2 & VQE & 0.75 &  71 &  37 \\
    PANSATZ \cite{meirom2023pansatz} & 2 & VQE & 0.7 &   40   & 0.7 \\
    Gate model: \Ry{} & 1 & VQ-SCI & 0.745&  71        &1.3195959(0.663096199) \\
    \rowcolor{gray!20}
    Freestyle (this work) & 1 & VQ-SCI & 0.745 &  0.22     &0.1745331950(26877544676) \\
    \bottomrule
    \end{tabular}
\end{table}

Our freestyle pulses did not \emph{always} reach \ca on the real device.
Potential reasons include: (a) statistical shot noise;
(b) leakage;  and (c)  measurements-induced errors.
Possible shot noise was excluded by trial -  we increased the number of shots up to $6\times10^5$ and observed no improvement.
As for leakage, it is typically associated with short, intense, and abrupt pulses leading to higher energy level occupation. However, a detailed analysis in the {Supplementary Material} shows that  in our specific case, where the solution is predominantly in the $\ket0$ state,
leakage is \emph{not} a major factor.
Lastly, we examined measurement errors.
We noted small deviation in $\oh Z$ measurements, whereas
$\oh X$  measurements, involving an extra Hadamard transformation, exhibited a significant error, see analysis in {Supplementary Material}.
We attribute inaccuracies in our calculations to this significant error.

\subsection{The LiH molecule} \label{sec:LiH}
We assessed the freestyle pulse method on multi-qubit circuits by computing the groundstate energy of the LiH molecule at its \SI{1.5}{\angstrom} equilibrium geometry, using the VQ-SCI scheme with 3 qubits \cite{yoffe2023qubitefficient}. In this geometry, the FCI energy of LiH is given by -7.882362 Ha, and the corresponding SCI energy, using 8 configurations, is -7.881566 Ha \cite{yoffe2023qubitefficient}.
Moving beyond the single qubit regime makes use of the 2-qubit control channels, which directly modulate the entanglement between qubits.
Since generating entanglement typically requires more time, and to facilitate our pulse search, we set a total duration of $\approx \SI{200}{\nano\second}$.

\paragraph{\textbf{Classical simulations}} \label{sec:LiH noisy simulations}
We fixed a frame of ten time bins, each of \SI{100}{\dt}, to a total duration of \SI{1000}{\dt}=\SI{222}{\nano\second}, per channel. Harboring the linear topology of \ibmqjakarta{}, led to 2 bidirectional 2-qubit channels, and 3 single qubit channels, amounting to 7 channels and a total of 140 real tunable parameters.
Optimizing this pulse setup with noiseless simulations reached a total groundsatate energy of -7.880692 Ha, corresponding to $\Delta_{FCI} = $\SI{1.67}{\milli\hartree} and $\Delta_{SCI} = $\SI{0.88}{\milli\hartree} \footnote{The accuracy of the pulse is dictated by the SCI energy difference since it operates within the VQ-SCI framework.}.
Noisy simulations of the same setup on \ibmqosaka{} resulted with $\Delta_{FCI} = $\SI{4.06}{\milli\hartree}.
The noiseless simulation outcome highlights the impact of the two-qubit control channels: accounting for only single-qubit drive channels in noiseless simulations yields a higher value of  $\Delta_{FCI}\!\!\approx$\SI{4}{\milli\hartree}.
The optimization process of this setup with 140 tunable parameters was notably time-consuming, requiring thousands of iterations for convergence. This is attributed to our choice of the gradient-free COBYLA  optimizer.
Employing gradient-based optimizers would necessitate an analog to the parameter shift rule  \cite{wierichs2022general,kyriienko2021generalized,kottmann2023evaluating}.

\paragraph{\textbf{Real-hardware}} \label{sec:real LiH hardware}

\begin{figure}[h!]
  \centering
  \includegraphics[width=0.6\linewidth]{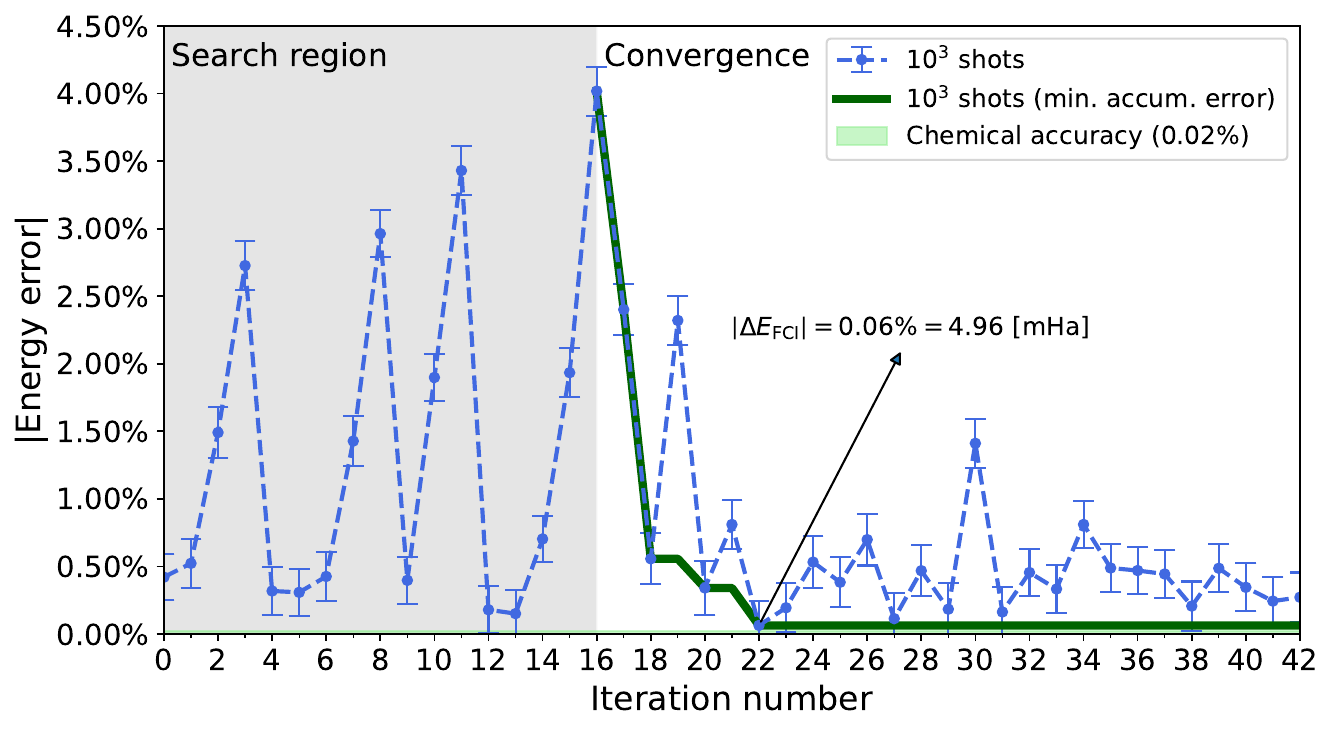}
  \caption{\small\textbf{Real-hardware} (\texttt{ibmq\textunderscore{}brisbane}) freestyle pulse results for \textbf{LiH} at equilibrium distance using $10^3$ shots. 
  Percentage error deviation from exact FCI energy shown in absolute value, per iteration. The dark green curve depicts the minimal accumulated error at each iteration.
  The light green area marks the \ca region.}
 \label{fig:LiH real-hardware}
\end{figure}

Our real-hardware executions, using $10^3$ shots, were performed upon the \texttt{ibmq\textunderscore{}brisbane} quantum computer, in which  \SI{1}{\dt}=\SI{0.5}{\nano\second}. To facilitate the executions we fixed a compact pulse setup, with 16 tunable real parameters. To that end, we enabled only uni-directional control pulses and fixed the following 4 time-bins sequence: all drive channels ($\SI{56}{\dt})\rightarrow$  first and second controls in turn ($\SI{152}{\dt} \ each)\rightarrow$ all drive channels ($\SI{56}{\dt}$), reaching a total duration of 416 dt.
Few tens of iterations were sufficient to recover the FCI energy to within \SI{4.96}{\milli\hartree}, 
as shown in \cref{fig:LiH real-hardware} (an experiment with $10^2$ shots achieved comparable results, not shown). The corresponding pulse (23rd iteration) is depicted in \cref{fig:LiH_pulse} in the {Supplementary Material}. 
A distinction between search and convergence region is made \footnote{When optimizing with COBYLA a problem with $p$ tunable parameters, the first $p$ iterations are dedicated to exploring the energy space}.

In comparison, a 3-qubit, depth-2 Ansatz, shown to be sufficient within VQ-SCI for \ca groundstate determination of the LiH molecule \cite{yoffe2023qubitefficient}, results in a pulse duration of $\approx\!\!\SI{1250}{\nano\second}$, six times longer than our freestyle pulse.
Moreover, the freestyle pulse attains much better accuracy on real-hardware, see \cref{table:LiH}.
Comparing further with previous pulse-based noisy-simulations of LiH, we note that in  \mcite{meitei2021gate} a shorter and more accurate pulse was found, albeit using a simplified pulse model, 
and in \mcite{meirom2023pansatz} a shorter pulse was reached, but with degraded accuracy. Finally, real-hardware pulse optimization in \mcite{liang2024napa} reached a pulse of comparable duration to ours, but with a much higher FCI energy deviation.

\begin{table}[]
    \scriptsize
    \centering
    \caption{
    \small \textbf{LiH} groundstate pulse calculations at the equilibrium interatomic distance (1.5 Ang). Per study, the table specifies: qubits number, utilized method, pulse duration, deviation from the FCI energy $(-7.88236 \ \text{Ha})$, and calculation type (noisy simulation or real-hardware).}
    \label{table:LiH}
    \begin{tabular}{
    @{}%
    l S[table-format=1] c
    S[table-format=4] S[table-format=3.2] 
    c
    @{}%
    }
    \toprule
    &{$\#q$}
    &{alg.}
    &{T [\unit{\nano\second}]}
    &{$\Delta_\mathrm{FCI}~\bqty{\unit{\milli\hartree}}$}
    &Calc. type
    \\
    \midrule
    Ctrl-VQE \cite{meitei2021gate} & 4 & VQE & 40 & 2.17 & Noisy sim.\\
    PANSATZ \cite{meirom2023pansatz} & 4 & VQE & 50 & 20 & Noisy sim. \\
    Gate model \cite{yoffe2023qubitefficient} & 3 & VQ-SCI & 1250 & 2.5     & Noisy sim.\\
    \rowcolor{gray!20}
    Freestyle (this work)  & 3 & VQ-SCI & 222 &    4.06     &  Noisy sim. \\
    \midrule
    \midrule
    NAPA \cite{liang2024napa} & 4 & VQE & 199 &  292 & RH \\
    Gate model \cite{yoffe2023qubitefficient} & 3 & VQ-SCI & 1250 & 81     & RH\\
    \rowcolor{gray!20}
    Freestyle (this work)  & 3 & VQ-SCI & 208 &    4.96     &  RH \\

    \bottomrule
    \end{tabular}
\end{table}

\section{Conclusions}
\label{sec:conclusions}

We introduced the freestyle pulse optimization scheme that provides full flexibility in pulse design. It differs from fixed-shape pulse methods in its ability to generate pulses of diverse, irregular, shapes. Additionally, unlike prior discrete-time pulse methods, our scheme accommodates two-qubit channels, thus enhancing its efficiency and compatibility with real quantum devices.

We demonstrated that shape-flexibility allows determining the groundstate energy of the \Htwo{} and LiH molecules to high accuracy while significantly reducing pulse duration. Notably, we achieved \ca{} for \Htwo{} groundstate calculation on an actual device using its minimal pulse duration of \SI{0.22}{\nano\second}. To the best of our knowledge, this is the shortest FCI-level accurate groundstate preparation, ever realized on real quantum hardware, closely reaching the theoretical quantum speed limit \cite{giovannetti2003quantum,giovannetti2004speed}.
Similarly, in calculating the LiH groundstate, we achieved unparalleled accuracy with pulse sequences significantly shorter than those generated by traditional gate-based circuits. 
These results underscore the potential advantages of delving into the hardware degrees of freedom. 
Finally, while our demonstrations focus on finding molecular groundstate energies, the approach is general and adaptable to various VQAs and hardware technologies.

To establish the practicality of the freestyle pulse optimization scheme it is required to: (a) demonstrate its scalability for multiple qubits circuits, involving numerous driving and control channels; and (b) evaluate its performance on additional VQAs.
These aspects are left for future research.

\section{Acknowledgments}
We acknowledge fruitful discussions with Prof. Emanuele Dalla Torre, Prof. Steven Frankel, and Dekel Meirom. We also acknowledge the use of IBM Quantum services for this work. The views expressed are those of the authors and do not reflect the official policy or position of IBM or the IBM Quantum team. NK and AM acknowledge the financial support of Elta and the Israel Innovation Authority through Meimad grant, number 77373.

\section{Code availability}
The underlying code for this study is available in
GitHub and can be accessed via this link \url{https://github.com/moroses/FreestylePulseOptimization}

\appendix
\section*{Supplementary Material}\label{sec:sup-material}
\subsection{The pulse Hamiltonian in superconducting qubits}
\label{sec:Hamilton}
In this section, we provide an in-depth view of \cref{eq:Hfull}.
The single-qubit Hamiltonian, denoted as the \enquote{Drive Hamiltonian}, is given, for the $k$'th qubit, by \cite{blais2021circuit}:
\begin{align}
\label{eq:H-drive}
\oh H^{(k)}_\mathrm{Drive}/\hbar=&
-\omega^z_{k}\oh\sigma^z_k+\Omega^\nd_kD_k\pqty{t}\oh\sigma_k^x,
\end{align}
where $\oh\sigma_k^\alpha$ are the Pauli operators, 
$\omega^z_{k}$ is the qubit's energy gap, $\Omega_k$ is the coupling constant, and,
\begin{align}
\label{eq:D}
D_k\pqty{t}=&\Re[e^{i\omega^d_{k}t}d_k\pqty{t}],
\end{align}
is the qubit's drive channel, where $\omega^d_{k}$ is the drive frequency and $d_k\pqty{t} \in \mathbb{C}$ is a user-defined time-dependent function, e.g.\ a Gaussian pulse.
To adhere to physical limitations, the drive function's magnitude obeys $\abs{d_k\pqty{t}}^2\le1$. 
The drive's frequency, $\omega^d_k$, is also user-controlled and can in principle be tuned. Here we use its default value, which is set to the qubit's energy gap, $\omega^z_k$.
The complex nature of $d_k\pqty{t}$ allows the implementation of the \Rz{} rotation virtually by applying a time-dependant phase to $d_k\pqty{t}$ \cite{mckay2017efficient}.

The interaction with each neighboring qubit $l$ is generated by the following cross-resonance Hamiltonian:
\begin{align}
\label{eq:H-CR}
\oh H^{(k,l)}_\mathrm{CR}/\hbar=&
J^\nd_{kl}\pqty{\oh\sigma_k^+\oh\sigma_l^-+\oh\sigma_k^-\oh\sigma_l^+},
\end{align}
where $J_{kl}$ is the coupling constant for qubit $k$ and $l$. User-defined control over the two-qubit interactions is gained via the control Hamiltonian
\begin{align}
\label{eq:H-control}
\oh H^{(k,l)}_\mathrm{Control}/\hbar=&
\Omega^\nd_k U^\nd_{k,l}\pqty{t}\oh\sigma_k^x,
\end{align}
with
\begin{align}
\label{eq:U}
U^\nd_{k,l}=&\Re[e^{i\pqty{\omega^d_{k}-\omega^d_{l}}t}u^\nd_{k,l}\pqty{t}],
\end{align}
where $u_{k,l}\pqty{t} \in \mathbb{C}$ is a user-defined time-dependent function for the control channel of qubit $k$ and $l$, in an analogy to $d_k(t)$ in the single qubit's drive channel, and the magnitude of the control function must obey $\abs{u_{k,l}\pqty{t}}^2\le1$.

\subsection{The \Htwo{} molecule in the VQ-SCI encoding}
\label{sec:H2-VQ-SCI}
Here we briefly outline the \Htwo{} molecule representation in the VQ-SCI approach, for more details see \cite{yoffe2023qubitefficient}.
The SCI Hamiltonian matrix of the \Htwo{} molecule, in the equilibrium interatomic distance of \SI{0.745}{\angstrom}, is given by:
\begin{align} \label{eq:H2_mat}
    \oh M_{CI} = &\pmqty{
               -1.8267 & 0.1814 \\
                0.1814 & -0.2596
            },
\end{align}
decomposed to the following sum of Pauli operators:
\begin{align}
    \label{eq:Pauli_decomposition_H2}
    \oh M_{CI} =& -1.0431\oh{I} -0.7835\oh{Z} + 0.1814\oh{X},
\end{align}
whose groundstate is given by the single qubit state: 
\begin{align}
    \label{eq:H2 groundstate}
    \ket{\Psi} = -0.9935\ket{0}+0.1135\ket{1},
\end{align}
corresponding to the lowest eigenvalue of \SI{-1.8474}{\hartree}.
After adding the nuclear repulsion energy (\SI{0.7103}{\hartree}), the total SCI groundstate energy amounts to 
\SI{-1.1371}{\hartree}, identical in this case to the FCI groundstate energy.

\subsection{The rotating wave approximation (RWA)}
\label{sec:RWA}
In single-qubit systems, it is often helpful to consider the
single qubit's drive channel under the rotating-wave approximation (RWA):
\begin{align}
\label{eq:q-rot}
\oh H^\mathrm{RWA}_\mathrm{Drive}/\hbar=\!\! -\delta^\nd_k\oh\sigma^z_k \!\!+\!\! \Omega^\nd_k \pqty{\Re[d^\nd_k\pqty{t}]\oh\sigma_k^x+\Im[d^\nd_k\pqty{t}]\oh\sigma_k^y},
\end{align}
where $\delta^\nd_k$ is the difference between the
frequencies of the qubit and the pump,
$\Omega_k$ is the coupling constant, and
 $\oh\sigma_k^\alpha$ is a Pauli operator
 with $\alpha \in \{x,y,z\}$ acting on qubit $k$.

 In this framework, the real and imaginary parts of $d(t)$ can be seen as rotations around the $x$ and $y$-axis, respectively; This explains the \Htwo{} results: 
 since a single $R_y$ gate alone is sufficient for accurately finding the groundstate of \Htwo{} \cite{yoffe2023qubitefficient}, the final pulse is expected to be mostly imaginary, as indeed observed in \cref{fig:convergence noisy simulation} in the main text.

\subsection{Leakage}
\label{sec:leakage}
Short and abrupt pulses are susceptible to leakage because the Fourier decomposition of such pulses yields a broad spectrum of frequency components, increasing the likelihood of off-target transitions \cite{gambetta2011analytic}.
To check if leakage affected our calculations, we analyzed our experimental raw measurements, namely the undiscriminated $I$ and $Q$ data points.

\begin{figure}[htb]
     \centering
     \begin{subfigure}[b]{0.48\linewidth}
         \centering
         \caption{}{}
         \label{fig:IQZ}
         \includegraphics[width=\linewidth]{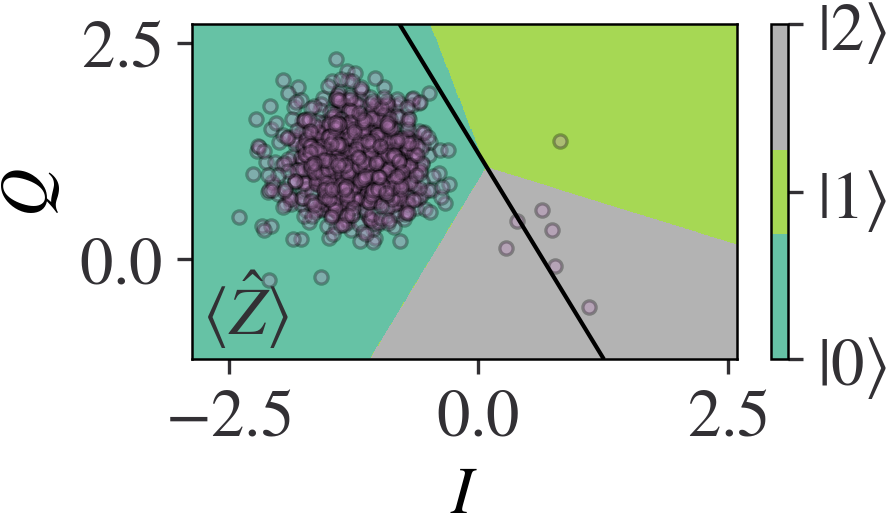}
     \end{subfigure}
     \hfill
     \begin{subfigure}[b]{0.48\linewidth}
         \centering
         \caption{}{}
         \label{fig:IQX}
         \includegraphics[width=\linewidth]{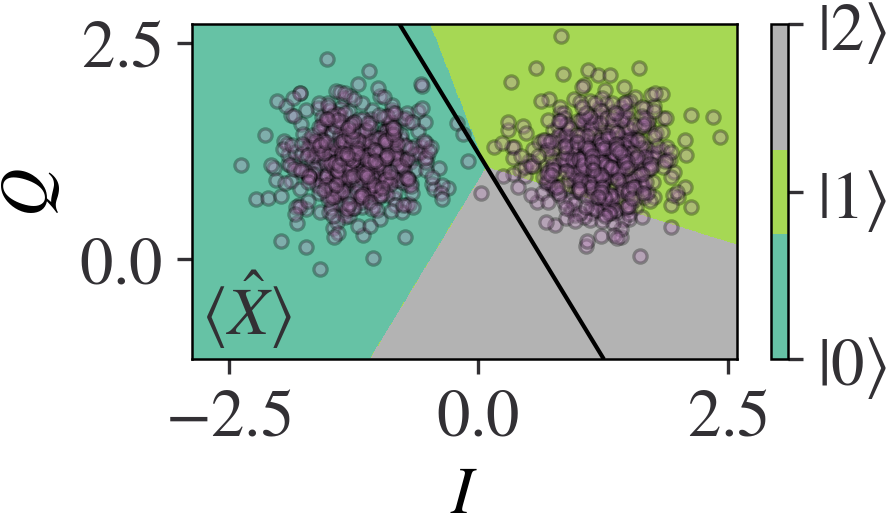}
     \end{subfigure}
     \caption{$IQ$ measurement points, in the $z$ and $x$ basis, classified to the $\ket{0}, \ket{1}$, and $\ket{2}$ states.
     The measurements were done in the final \Htwo{} VQ-SCI iteration.
     The $I$ and $Q$ data points are normalized by a factor of $10^8$.
     Each dot corresponds to a single measured shot on the quantum computer.
     The solid line represents the classification to $\ket0$ and $\ket1$.}
     \label{fig:IQ_classificaltion}
\end{figure}

We employed IBM's standard calibration process \cite{0-1-2}, to discriminate the $I$ and $Q$ data points to different quantum states, namely, $\ket0$, $\ket1$, and $\ket2$ to identify leakage.
This process is composed of standard frequency spectroscopy for identifying transition frequencies and Rabi spectroscopy for calibrating transition pulse amplitudes.
Now we can define regions in the $I$ and $Q$ planes to represent the different quantum states.
To that end, we used the linear discriminant analysis (LDA) class in ``scikit-learn" \cite{scikit-learn}. 
Then, we classified the $I$ and $Q$ measurements acquired at each iteration to these three states, as depicted in \cref{fig:IQ_classificaltion} for the final VQ-SCI iteration in the groundstate calculation of \Htwo{}; the solid line corresponds to $\{\ket{0},\ket{1}\}$ classification.
It is seen that in the $\oh Z$-measurement, see \cref{fig:IQZ}, the leakage is negligible, whereas in the $\oh X$-measurements, see \cref{fig:IQX}, more $IQ$ points are found in the $\ket{2}$ state region.
Yet, their vast majority can be safely associated with the $\ket{1}$ state (one may consider using a different classifier, but LDA seems to be sufficient, given the minor leakage). 
This hints that the effect of leakage in our calculation is negligible and is not the main source of error in real devices.

\begin{figure}[htb]
  \centering
  \includegraphics[width=0.7\linewidth]{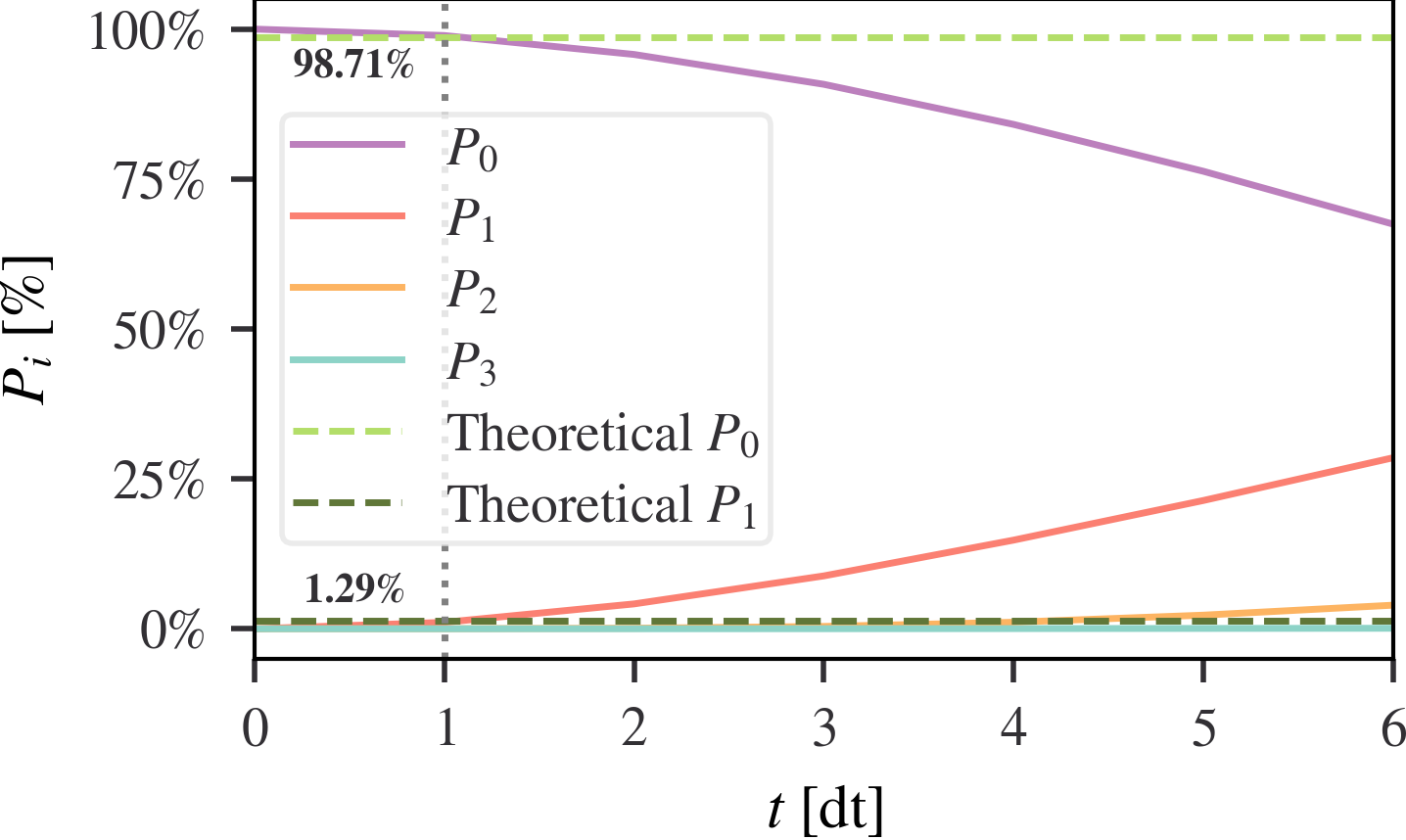}
  \caption{
  Simulating the evolution of a 4-level system, starting from the $\ket0$ state, under a constant pulse of maximal amplitude, as a function of time.
  The simulation is noisy, based on Jakarta's topology and noise model, and assuming \enquote{right-padding}, see text.
  This simulation supports the finding that a single pulse of duration \SI{1}{\dt} (\SI{.222}{\nano\second}) recovers the groundstate of \Htwo{} to good accuracy and that the leakage is minimal.
  }
  \label{fig:4-levels_system}
\end{figure}

To further support this claim, we simulated the dynamics of a 4-level system, undergoing a single pulse of \textit{maximal amplitude}.
\Cref{fig:4-levels_system} shows the probability of each state as a function of time, starting from the $\ket{0}$ state.
It is seen that the probability of measuring the $\ket{0}$ state (purple) drops with time, whereas the probability of measuring the $\ket{1}$ state (red) increases with time, as expected.
As a sanity check, \cref{fig:4-levels_system} depicts in horizontal dashed lines the expected probabilities of the $\ket{0}$ (light green) and $\ket{1}$ (dark green) states in the \Htwo{} FCI groundstate, as given in \cref{eq:H2 groundstate}.
It is seen that a maximal amplitude pulse attains the required state after \SI{1}{\dt}, in good corresponding to our classical simulation results, see \cref{fig:convergence noisy simulation}.

As for leakage, it is seen that the probability of measuring the $\ket{2}$ state (orange), increases very slowly with time, reaching only \SI{\sim2}{\percent} leakage to the $\ket{2}$ state after \SI{5}{\dt} and that the probability of measuring the $\ket{3}$ state (cyan) remains negligible even after \SI{6}{\dt}. This implies that even at maximum amplitude intensity a steady pulse would experience very little leakage after \SI{1}{\dt}.
Moreover, with more typical amplitudes, of up to
\SI{30}{\percent} of the maximal amplitude, much less leakage is expected.
For example, for the best iteration from \cref{fig:real-hardware} in the main text which has an amplitude of:

\begin{align}
d\pqty{0<t\leq\SI{0.22}{\nano\second}}=&\complexnum{-0.070-0.266i},
\end{align}
the expected leakage is practically zero, with merely $2.45 \cdot 10^{-5}\,\%$.

This analysis underscores the reliability of our experimental outcomes, demonstrating that leakage, even under conditions of short and strong pulses, is minimal in our particular case, where the precise solution resides primarily in the $\ket{0}$ state (see \cref{eq:H2 groundstate}).

\subsection{Measurement error}
\label{sec:ME}
In the process of measuring the components of the Hamiltonian, see \cref{eq:Pauli_decomposition_H2}, the identity term $\oh I$ is straightforwardly added without physical measurement, the $\oh Z$ term is measured in the standard computational basis, and the $\oh X$ term requires a simple basis transformation using the Hadamard gate.

\begin{figure}[htb]
  \centering
  \includegraphics[width=0.6\linewidth]{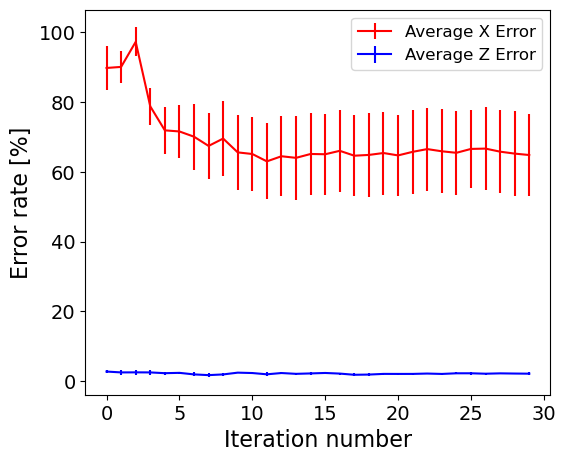}
  \caption{Error deviation of $\hat{X}$-$\hat{Z}$ measurement from the expected values of the FCI groundstate on the equilibrium of \Htwo{}. This figure illustrates an absolute error analysis for each iteration of the average of the 4 freestyle executions on real-hardware, shown in blue in \cref{fig:real-hardware}. The measurement errors for the $\hat{X}$ and $\hat{Z}$ operators are depicted by red and blue lines, respectively. The error bars represent the standard error of the mean (SEM), based on the outcomes of measurement shots.}
  \label{fig:error_diff_base}
\end{figure}

 To pinpoint the source of errors in our results, we conducted a thorough analysis focusing on measurement errors. 
 Our strategy involved leveraging the known groundstate of the \Htwo{} molecule, as determined by the VQ-SCI method (see \cref{eq:H2 groundstate}).
 This allowed us to calculate the precise expectation values.
 The results of this analysis are presented in \cref{fig:error_diff_base}.
 This figure includes multiple iterations of our experiment on the \ibmqjakarta{} quantum device, all conducted using a single time step (\unit\dt) and the same number of shots ($10^5$). These results are based on the same experiments presented in the blue curve in \cref{fig:real-hardware}.
 The figure focuses on the measurement errors of the expectation values of the $\oh Z$ and $\oh X$ operators.
 Each point represents the average of 4 different freestyle pulse experiments. Our analysis revealed a notable difference in the measurement accuracy between the two operators.
 Specifically, the expectation value for the $\oh X$ operator showed a significant deviation,
 saturating at \SI{\approx60}{\percent} measurement error, whereas the $\oh Z$ operator measurements were more accurate, with deviations at \SI{\approx5}{\percent}.

 Note that in the specific Hamiltonian of \cref{eq:Pauli_decomposition_H2}, the contribution of the $\oh Z$ term to the overall measurement is approximately four times greater than that of the $\oh X$ term, and is hence much more dominant. In addition, the two terms are of opposite signs, giving rise to occasional observed measurement error cancellation.

\subsection{The LiH molecule in SCI encoding using 8 Slater determinants}
\label{sec:LiH-SCI}
The {LiH} on the equilibrium of \SI{1.5}{\angstrom} can be described by the following SCI Hamiltonian matrix \cite{yoffe2023qubitefficient}:
 \begin{align*}
 \pqty{
 \begin{smallmatrix}
 -8.922 & 0.123 & 0.033 & 0.033 & 0.000 & 0.000 & 0.012 & 0.024 \\
 0.123 & -7.819 & -0.025 & -0.025 & -0.031 & -0.031 & 0.026 & 0.019 \\
 0.033 & -0.025 & -8.178 & 0.026 & -0.008 & 0.142 & -0.104 & 0.014 \\
 0.033 & -0.025 & 0.026 & -8.178 & 0.142 & -0.008 & -0.104 & 0.014 \\
 0.000 & -0.031 & -0.008 & 0.142 & -8.764 & 0.012 & 0.053 & 0.019 \\
 0.000 & -0.031 & 0.142 & -0.008 & 0.012 & -8.764 & 0.053 & 0.019 \\
 0.012 & 0.026 & -0.104 & -0.104 & 0.053 & 0.053 & -8.226 & 0.041 \\
 0.024 & 0.019 & 0.014 & 0.014 & 0.019 & 0.019 & 0.041 & -8.252 \\
 \end{smallmatrix}
 }
 \end{align*}
which is decomposed to 36 Pauli operators and corresponds to the groundstate energy of \SI{-8.93992}{\hartree}. Together with the nuclear repulsion energy of \SI{1.05835}{\hartree}, the total SCI groundstate energy amounts to  \SI{-7.88157}{\hartree}, which is \SI{0.79}{\milli \hartree} away from the exact FCI total energy,  given by \SI{-7.88236}{\hartree}.

\subsection{Quantum speed limit}
\label{sec:QSL}
{
\sisetup{
round-mode=places,
round-precision=5,
}
Quantum speed limit (QSL) is the minimal time required for evolving a given initial state $\ket{\psi_0}$ to a final state $\ket{\psi_f}$ \cite{giovannetti2003quantum}.
While it is not a rigorous bound that can not be violated (see \cite{atia2017fast} for example), it gives an order of magnitude benchmark for the required time evolution.

Here we derive the QSL for the groundstate energy calculation of   \Htwo{}, starting from the $\ket{0}$ state. To that end, we follow the procedure described in \mcitep{giovannetti2003quantum,giovannetti2004speed} for calculating the QSL for transition between non-orthogonal states, given by:

\begin{align}
    \label{eq:qsl:def}
    \tau_\mathrm{QSL}\equiv&\max\qty(\alpha\pqty{\varepsilon}\frac{\hbar\pi}{2\bar{E}},\beta\pqty{\varepsilon}\frac{\hbar\pi}{2\Delta \bar{E}}),
\end{align}
where
\begin{align}
    \label{eq:qsl:ave_e}
    \bar{E}=&\frac1{T}\int\limits_{0}^TE\pqty{t}\dd{t}
\end{align}
is the time average energy of the system, and
\begin{align}
    \label{eq:qsl:ave_var_e}
    \Delta\bar{E}=&\frac1T\int\limits_{0}^{T}\sqrt{
    \mel{
    \psi\pqty{t}
    }
    {
    \pqty{\oh H\pqty{t}-E\pqty{t}}^2
    }
    {\psi\pqty{t}
    }
    }
    \dd{t}
\end{align}
is the time average energy variance of the system;
where $E\pqty{t}\equiv\mel{\psi\pqty{t}}{\oh H\pqty{t}}{\psi{\pqty{t}}}$ is the energy of a given state at the time $t$ and where the Hamiltonian $\oh H\pqty{t}$ is shifted by the groundstate energy such that the lowest energy is 0.
The $\alpha\pqty{\varepsilon}$ and $\beta\pqty{\varepsilon}$ are functions of the distance $\epsilon$ between the initial and the final state, which is given by:
\begin{align}
    \label{eq:qsl:epsilon}
    \varepsilon
    \equiv&
    \abs{\braket{\psi_0}{\psi_f}}^2.
\end{align}
The function $\alpha\pqty{\varepsilon}$ has no analytical expression, but it can be approximated by $\alpha\pqty{\varepsilon}\simeq\beta^2\pqty{\varepsilon}$ (see detailed derivation in \mcitep{giovannetti2003quantum,giovannetti2004speed}), where the $\beta\pqty{\varepsilon}$ function is given by:
\begin{align}
    \label{eq:qsl:beta}
    \beta\pqty{\varepsilon}=&\frac2\pi \arccos(\sqrt{\varepsilon}).
\end{align}

In the superconducting setting, there are two candidate frames for calculating the QSL, the lab frame:
\begin{align}
    \oh H^\mathrm{lab}\pqty{t=0}/\hbar=&-\omega_z\oh\sigma_z,
    \\ \nonumber
    \oh H^\mathrm{lab}\pqty{t=\unit\dt}/\hbar=&-\omega_z\oh\sigma_z
    +\Re[e^{-i\omega_z \unit\dt}d\pqty{t=\unit\dt}]\Omega\oh\sigma_x,
\end{align}
and the rotating-wave-approximation (RWA) frame:
\begin{align}
    \oh H^\mathrm{rwa}\pqty{t=0}/\hbar=&-\delta_z\oh\sigma_z,
    \\ \nonumber
    \oh H^\mathrm{rwa}\pqty{t=\unit\dt}/\hbar=&-\delta_z\oh\sigma_z
    +\Re[d\pqty{t=\unit\dt}]\Omega\oh\sigma_x
    \\& \nonumber
    \quad \quad +\Im[d\pqty{t=\unit\dt}]\Omega\oh\sigma_y
    .
\end{align}

For each frame, we considered two possible final states: the theoretical groundstate, given in \cref{eq:H2 groundstate}, and the best state we reached experimentally (corresponding to the cyan curve in \cref{fig:real-hardware}). We calculated the experimental state by simulating the unitary evolution the qubit experienced, in both the lab and the RWA frames, based on the Hamiltonian parameters used in the experiment, see \cref{tab:parameters}. This resulted in the following experimental state for the lab frame:

{
\sisetup{
round-mode=places,
round-precision=4,
}
\begin{align}
    \ket{\psi^\mathrm{lab}_e}\equiv&\ket{\psi\pqty{t=\unit\dt}}=
    U^\mathrm{lab}\ket0\\=& \nonumber
    \pqty{\complexnum{0.999918-0.00011886i}}\ket0
    +\pqty{\complexnum{-0.00724387-0.0105343i}}\ket1,
\end{align}
and for the RWA frame:
\begin{align}
    \ket{\psi^\mathrm{rwa}_e}\equiv&\ket{\psi\pqty{t=\unit\dt}}=
    U^\mathrm{rwa}\ket0\\=& \nonumber
    \pqty{\complexnum{0.98916-0.10355i}}\ket0
    +\pqty{\complexnum{-0.0973245+0.0369881i}}\ket1,
\end{align}
}

\begin{table}[htb]
    \centering
    \caption{Parameters used to calculate the QSL.}
    \label{tab:parameters}
    \begin{tabular}{
    @{}
    r
    @{$=$}
    S[table-format=+3.3, table-auto-round]
    l
    @{}
    }
    \toprule
    \unit\dt&.222&\unit{\nano\second}
    \\
    $\delta_z$&0&\unit{\pibar\giga\hertz}
    \\
    $\Omega$&271.3740963766729&\unit{\pibar\mega\hertz}
    \\
    $\Re[d\pqty{t}]$&-0.07047161691763452
    \\
    $\Im[d\pqty{t}]$&-0.26610396687855536
    \\
    \bottomrule
    \end{tabular}
\end{table}

\begin{table*}[htb]
    \centering
    \caption{
    The various components of the QSL calculation according to \cref{eq:qsl:def}.
    }
    \label{tab:qsl}
    \sisetup{
    table-auto-round,
    }
    \rowcolors{2}{gray!25}{white}
    \begin{tabular}{
    @{}
    *{2}{l}
    S[table-format=2.2, exponent-mode=fixed, fixed-exponent=-2]
    *{2}{S[table-format=1.2, drop-exponent=false, exponent-mode=fixed, fixed-exponent=-2]}
    S[table-format=2.2]
    *{1}{S[table-format=3.2]}
    S[table-format=1.3]
    S[table-format=1.3]
    S[table-format=1.2]
    @{}
    }
    \toprule
    \\
    {Frame}&
    {Final state}&
    {$\varepsilon~\bqty{\times10^{-2}}$}&
    {$\beta~\bqty{\times10^{-2}}$}&
    {$\alpha~\bqty{\times10^{-2}}$}&
    {$\bar{E}~\bqty{\unit{\ebar\mega\hertz}}$}&
    {$\Delta\bar{E}~\bqty{\unit{\ebar\mega\hertz}}$}&
    \multicolumn{2}{c}{{$\tau_\mathrm{QSL}~\bqty{\unit{\nano\second}}$}}&
    {$\tau^\mathrm{max}_\mathrm{QSL}~\bqty{\mathrm{dt}}$}
    \\
    \cmidrule(lr){8-9}
    &&&&&&&{$\alpha\pqty{\varepsilon}\hbar\pi/\pqty{2\bar{E}}$}&
    {$\beta\pqty{\varepsilon}\hbar\pi/\pqty{2\Delta\bar{E}}$}
    \\
    \midrule
    Lab&theoretical&.987042&.0726252&.00527442&56.106&545.655&.023502&.0332743&.149734
    \\
    Lab&
    experimental&
    .999837& .00813915& 6.62458e-5& .68265& 60.502& .0242605& .0336318& .151343
    \\
    RWA&
    theoretical&
    .987042& .0726252& .00527442& 79.0104& 74.5757& .016689& .243461& 1.09558
    
    \\
    RWA&
    experimental&
    .98916& .0664028& .00440933& 74.7031& 74.7031& .0147562& .222222& 1.0

    \\
         \bottomrule
    \end{tabular}
\end{table*}
}

{
\sisetup{
round-mode=places,
round-precision=3,
}

\Cref{tab:qsl} summarizes the full calculations of the QSL for any combination of the frame and state, showing that the QSLs in the lab frame are:
\begin{align}
    \tau^\mathrm{lab}_\mathrm{QSL,t}=\SI{.0332743}{\nano\second}, \quad \tau^\mathrm{lab}_\mathrm{QSL,e}=\SI{.0336318}{\nano\second}
\end{align}
whereas in the RWA, the QSLs are:
\begin{align}
    \tau^\mathrm{rwa}_\mathrm{QSL,t}=\SI{.243461}{\nano\second}, \quad
    \tau^\mathrm{rwa}_\mathrm{QSL,e}=\SI{.222222}{\nano\second}.
\end{align}

It is seen that whether we use the experimental final state or the theoretical final state, the obtained QSLs are close in both frames.
However, there is a difference of an order of magnitude ($\tau^\mathrm{rwa}_\mathrm{QSL}\sim10\tau^\mathrm{lab}_\mathrm{QSL}$) between the two frames.
In the RWA frame, all the fast oscillating terms are dropped, thus providing a 
stricter bound for the QSL. 
As we can see, in all limits our results are in agreement with the QSL. Moreover, under the specific parameters of our experimental pulse, the QSL of the RWA for our resulting experimental state is exactly $1 dt$. This indicates that within the device constraints, our experiment recovered the theoretical QSL.

To estimate the shortest possible time evolution on the same device (\ibmqjakarta{}), we repeated the calculation with the same experimental parameters, but this time with a maximal amplitude. Our calculation showed that for such a setup the QSLs can be reduced to:
\begin{align}
    \tau^\mathrm{rwa}_\mathrm{QSL^*,t}=\SI{.067}{\nano\second}, \quad
    \tau^\mathrm{rwa}_\mathrm{QSL^*,e}=\SI{.061}{\nano\second},
\end{align}
so that in principle, if the minimal time bin $dt$ could be relaxed, the exact FCI groundstate for the \Htwo{} molecule could be reached via a pulse of \SI{.067}{\nano\second}. Of course, experimental dispersion and additional qubit/cavity modes will limit the realistic bandwidth of an actual pulse, making this extremely short driving unrealistic, although an important theoretical order-of-magnitude in understanding many-body control limitations.

\subsection{The LiH pulse}
\label{sec:LiH-pulse}
\Cref{fig:linear_topology} shows the linear topology of the three qubits from the \texttt{ibmq\textunderscore{}brisbane} quantum device that we used. The setup involved a single complex drive channel for each qubit, in addition to two complex directed 2-qubit control channels, resulting in a total of five pulse channels.
\Cref{fig:LiH_pulse} depicts the most accurate 3-qubit 5 channels pulse generating the LiH groundstate attained on real-hardware, corresponding to the 23'rd iteration in the convergence process shown in \cref{fig:LiH real-hardware}, see main text.

\begin{figure}[htb]
  \centering
  \begin{subfigure}{\linewidth}
    \caption{}{}
    \label{fig:linear_topology}
    \includegraphics[width=0.3\linewidth]{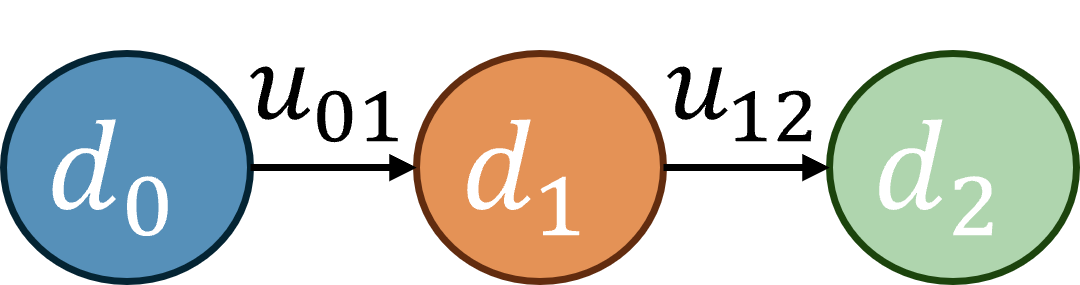}
  \end{subfigure}
  \\
  \begin{subfigure}{\linewidth}
    \centering
    \caption{}{}
    \label{fig:LiH_pulse}
    \includegraphics[width=\linewidth]{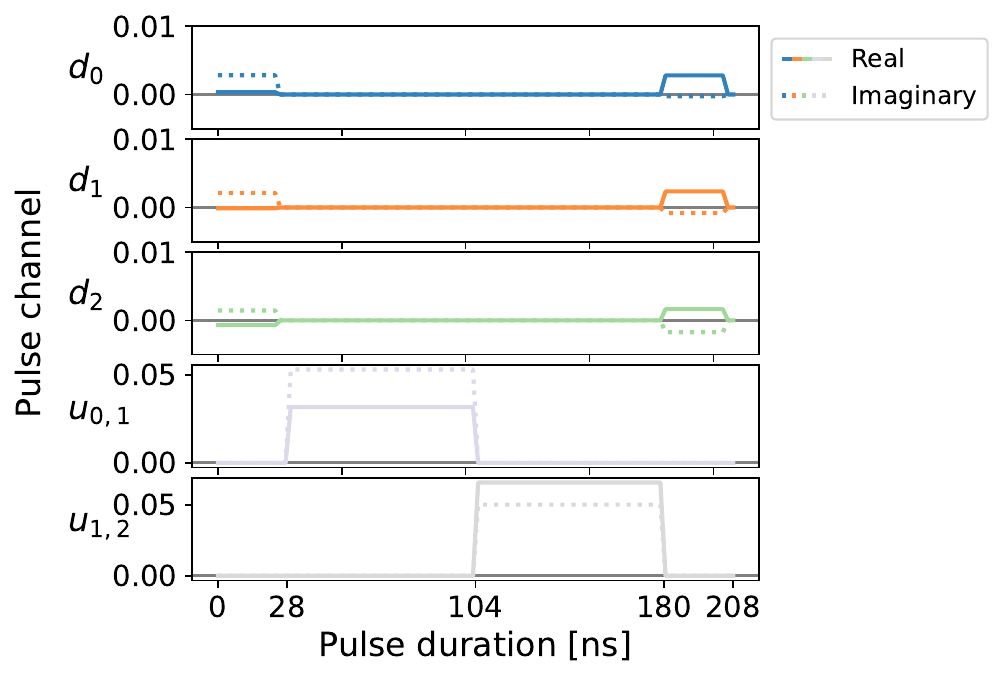}
  \end{subfigure}
  \caption{\textbf{(\subref{fig:linear_topology}) Top:} An illustration of the 3-qubit topology featuring three drive channels, one assigned to each qubit $i$, and two control channels, denoted as $u_{k,l}$, where $k$ is the controlling qubit and $l$ is the target qubit; \textbf{(\subref{fig:LiH_pulse}) Bottom:} The real-hardware pulse schedule that most closely approximated the FCI groundstate energy for LiH (found at the 23rd iteration, see \cref{fig:LiH real-hardware} in the main text). The $y$-axis displays the optimized pulse amplitudes, while the $x$-axis outlines the timeline of the pulse. Solid and dotted lines indicate real and imaginary components of the pulse amplitudes, respectively. Here, $d_i$ denotes the three single-qubit drive channels, and $u_{k,l}$ signifies the two two-qubit control channels. }
  \label{fig:LiH_stuff}
\end{figure}

\bibliography{apssamp}

\end{document}